\documentclass[aps,prl,reprint,superscriptaddress,twocolumn]{revtex4-1}

\usepackage{graphicx}
\usepackage{amsmath}
\usepackage{amssymb}
\usepackage{units}
\usepackage{color}

\begin{document}

\title{Flavor Ratio of Astrophysical Neutrinos above 35 TeV in IceCube}

\affiliation{III. Physikalisches Institut, RWTH Aachen University, D-52056 Aachen, Germany}
\affiliation{School of Chemistry \& Physics, University of Adelaide, Adelaide SA, 5005 Australia}
\affiliation{Dept.~of Physics and Astronomy, University of Alaska Anchorage, 3211 Providence Dr., Anchorage, AK 99508, USA}
\affiliation{CTSPS, Clark-Atlanta University, Atlanta, GA 30314, USA}
\affiliation{School of Physics and Center for Relativistic Astrophysics, Georgia Institute of Technology, Atlanta, GA 30332, USA}
\affiliation{Dept.~of Physics, Southern University, Baton Rouge, LA 70813, USA}
\affiliation{Dept.~of Physics, University of California, Berkeley, CA 94720, USA}
\affiliation{Lawrence Berkeley National Laboratory, Berkeley, CA 94720, USA}
\affiliation{Institut f\"ur Physik, Humboldt-Universit\"at zu Berlin, D-12489 Berlin, Germany}
\affiliation{Fakult\"at f\"ur Physik \& Astronomie, Ruhr-Universit\"at Bochum, D-44780 Bochum, Germany}
\affiliation{Physikalisches Institut, Universit\"at Bonn, Nussallee 12, D-53115 Bonn, Germany}
\affiliation{Universit\'e Libre de Bruxelles, Science Faculty CP230, B-1050 Brussels, Belgium}
\affiliation{Vrije Universiteit Brussel, Dienst ELEM, B-1050 Brussels, Belgium}
\affiliation{Dept.~of Physics, Chiba University, Chiba 263-8522, Japan}
\affiliation{Dept.~of Physics and Astronomy, University of Canterbury, Private Bag 4800, Christchurch, New Zealand}
\affiliation{Dept.~of Physics, University of Maryland, College Park, MD 20742, USA}
\affiliation{Dept.~of Physics and Center for Cosmology and Astro-Particle Physics, Ohio State University, Columbus, OH 43210, USA}
\affiliation{Dept.~of Astronomy, Ohio State University, Columbus, OH 43210, USA}
\affiliation{Niels Bohr Institute, University of Copenhagen, DK-2100 Copenhagen, Denmark}
\affiliation{Dept.~of Physics, TU Dortmund University, D-44221 Dortmund, Germany}
\affiliation{Dept.~of Physics and Astronomy, Michigan State University, East Lansing, MI 48824, USA}
\affiliation{Dept.~of Physics, University of Alberta, Edmonton, Alberta, Canada T6G 2E1}
\affiliation{Erlangen Centre for Astroparticle Physics, Friedrich-Alexander-Universit\"at Erlangen-N\"urnberg, D-91058 Erlangen, Germany}
\affiliation{D\'epartement de physique nucl\'eaire et corpusculaire, Universit\'e de Gen\`eve, CH-1211 Gen\`eve, Switzerland}
\affiliation{Dept.~of Physics and Astronomy, University of Gent, B-9000 Gent, Belgium}
\affiliation{Dept.~of Physics and Astronomy, University of California, Irvine, CA 92697, USA}
\affiliation{Dept.~of Physics and Astronomy, University of Kansas, Lawrence, KS 66045, USA}
\affiliation{Dept.~of Astronomy, University of Wisconsin, Madison, WI 53706, USA}
\affiliation{Dept.~of Physics and Wisconsin IceCube Particle Astrophysics Center, University of Wisconsin, Madison, WI 53706, USA}
\affiliation{Institute of Physics, University of Mainz, Staudinger Weg 7, D-55099 Mainz, Germany}
\affiliation{Universit\'e de Mons, 7000 Mons, Belgium}
\affiliation{Technische Universit\"at M\"unchen, D-85748 Garching, Germany}
\affiliation{Bartol Research Institute and Dept.~of Physics and Astronomy, University of Delaware, Newark, DE 19716, USA}
\affiliation{Department of Physics, Yale University, New Haven, CT 06520, USA}
\affiliation{Dept.~of Physics, University of Oxford, 1 Keble Road, Oxford OX1 3NP, UK}
\affiliation{Dept.~of Physics, Drexel University, 3141 Chestnut Street, Philadelphia, PA 19104, USA}
\affiliation{Physics Department, South Dakota School of Mines and Technology, Rapid City, SD 57701, USA}
\affiliation{Dept.~of Physics, University of Wisconsin, River Falls, WI 54022, USA}
\affiliation{Oskar Klein Centre and Dept.~of Physics, Stockholm University, SE-10691 Stockholm, Sweden}
\affiliation{Dept.~of Physics and Astronomy, Stony Brook University, Stony Brook, NY 11794-3800, USA}
\affiliation{Dept.~of Physics, Sungkyunkwan University, Suwon 440-746, Korea}
\affiliation{Dept.~of Physics, University of Toronto, Toronto, Ontario, Canada, M5S 1A7}
\affiliation{Dept.~of Physics and Astronomy, University of Alabama, Tuscaloosa, AL 35487, USA}
\affiliation{Dept.~of Astronomy and Astrophysics, Pennsylvania State University, University Park, PA 16802, USA}
\affiliation{Dept.~of Physics, Pennsylvania State University, University Park, PA 16802, USA}
\affiliation{Dept.~of Physics and Astronomy, Uppsala University, Box 516, S-75120 Uppsala, Sweden}
\affiliation{Dept.~of Physics, University of Wuppertal, D-42119 Wuppertal, Germany}
\affiliation{DESY, D-15735 Zeuthen, Germany}

\author{M.~G.~Aartsen}
\affiliation{School of Chemistry \& Physics, University of Adelaide, Adelaide SA, 5005 Australia}
\author{M.~Ackermann}
\affiliation{DESY, D-15735 Zeuthen, Germany}
\author{J.~Adams}
\affiliation{Dept.~of Physics and Astronomy, University of Canterbury, Private Bag 4800, Christchurch, New Zealand}
\author{J.~A.~Aguilar}
\affiliation{Universit\'e Libre de Bruxelles, Science Faculty CP230, B-1050 Brussels, Belgium}
\author{M.~Ahlers}
\affiliation{Dept.~of Physics and Wisconsin IceCube Particle Astrophysics Center, University of Wisconsin, Madison, WI 53706, USA}
\author{M.~Ahrens}
\affiliation{Oskar Klein Centre and Dept.~of Physics, Stockholm University, SE-10691 Stockholm, Sweden}
\author{D.~Altmann}
\affiliation{Erlangen Centre for Astroparticle Physics, Friedrich-Alexander-Universit\"at Erlangen-N\"urnberg, D-91058 Erlangen, Germany}
\author{T.~Anderson}
\affiliation{Dept.~of Physics, Pennsylvania State University, University Park, PA 16802, USA}
\author{C.~Arguelles}
\affiliation{Dept.~of Physics and Wisconsin IceCube Particle Astrophysics Center, University of Wisconsin, Madison, WI 53706, USA}
\author{T.~C.~Arlen}
\affiliation{Dept.~of Physics, Pennsylvania State University, University Park, PA 16802, USA}
% ...

\date{\today}

\collaboration{IceCube Collaboration}
\noaffiliation

\begin{abstract}
A diffuse flux of astrophysical neutrinos above $100\,\mathrm{TeV}$ has been observed at the IceCube Neutrino Observatory.  Here we extend this analysis to probe the astrophysical flux down to $35\,\mathrm{TeV}$ and analyze its flavor composition by classifying events as showers or tracks.  Taking advantage of lower atmospheric backgrounds for shower-like events, we obtain a shower-biased sample containing 129 showers and 8 tracks collected in three years from 2010 to 2013.  We demonstrate consistency with the $(f_e:f_{\mu}:f_\tau)_\oplus\approx(1:1:1)_\oplus$ flavor ratio at Earth commonly expected from the averaged oscillations of neutrinos produced by pion decay in distant astrophysical sources.  Limits are placed on non-standard flavor compositions that cannot be produced by averaged neutrino oscillations but could arise in exotic physics scenarios.  A maximally track-like composition of $(0:1:0)_\oplus$ is excluded at $3.3\sigma$, and a purely shower-like composition of $(1:0:0)_\oplus$ is excluded at $2.3\sigma$.
\end{abstract}

\maketitle

\textit{Introduction---}Traveling virtually unimpeded through matter, radiation, and magnetic fields, astrophysical neutrinos probe otherwise inaccessible regions of the high-energy universe.  If produced from cosmic rays interacting with gas and radiation at their sources, they convey unique information about astrophysical particle accelerators in their direction, energy, and flavor \cite{bz,1995PhR...258..173G,2000ARNPS..50..679L,2002RPPh...65.1025H,2008PhR...458..173B}.  Though no individual sources of $\mathrm{TeV}$ cosmic neutrinos have yet been found, a diffuse flux was observed by the IceCube Neutrino Observatory above $100\;\mathrm{TeV}$ in three years of data \cite{hese_paper,hese2}.  Here this work is expanded to observe the diffuse astrophysical neutrino flux down to $35\;\mathrm{TeV}$ and derive constraints on its flavor composition.

Astrophysical neutrinos are expected from the decay of secondary particles such as pions, kaons, muons, and neutrons produced in cosmic ray interactions. In the model of diffusive shock acceleration, the differential energy spectrum of injected cosmic rays follows a power law $\propto E^{-\gamma}$ with $\gamma\sim2$ \cite{0034-4885-46-8-002,Blandford19871}.  Though this spectrum may be modified by propagation in cosmic magnetic fields en route to Earth \cite{crprop}, neutrinos produced at the source retain the same spectral index $\gamma$ as the injected cosmic ray spectrum, provided the environment is sparse enough to allow particles to decay rather than interact.  In the most commonly considered scenario, the decay of pions and their daughter muons dominate the neutrino flux, resulting in a flavor ratio of $(f_{e}:f_{\mu}:f_{\tau})_S=(1:2:0)_S$ at source \cite{2006MPLA...21.1049A,nu_flavorratio}. However, the composition could vary from $(0:1:0)_S$ to $(1:0:0)_S$ under a multitude of scenarios including muon energy loss in high matter density or magnetic fields \cite{PhysRevD.58.123005,PhysRevLett.95.181101,PhysRevD.74.063009,2007PhRvD..75l3005L}, muon acceleration \cite{0004-637X-779-2-106}, and neutron decay \cite{Anchordoqui200442}.

As first noted in \cite{Learned:1994wg}, neutrino oscillations, averaged by propagation over astronomical distances, transform the flavor ratio according to the PMNS mixing matrix \cite{PhysRevD.75.113004,Esmaili2009197,2009PhRvD..80k3006C}.  Taking global best-fit mixing parameters \cite{nufit}, a flavor ratio at Earth of $(f_{e}:f_{\mu}:f_{\tau})_\oplus = (0.93:1.05:1.02)_{\oplus}\approx(1:1:1)_{\oplus}$ is expected for a $(1:2:0)_S$ source composition, a result of the near tribimaximal form of the PMNS matrix \cite{2002PhLB..530..167H}.  For a composition at sources varying from $(0:1:0)_S$ to $(1:0:0)_S$, the composition at Earth varies linearly from $(0.6:1.3:1.1)_\oplus$ to $(1.6:0.6:0.8)_\oplus$.  Though expected to be negligible \cite{2006MPLA...21.1049A}, even a large $\nu_{\tau}$ contribution at sources causes only a small deviation from this range.  Because of this limited variation in the flavor ratio at Earth for all possible source compositions, the observation of a ratio inconsistent with these expectations would signal new physics in the neutrino sector, such as neutrino decay \cite{PhysRevLett.90.181301,1475-7516-2012-10-020}, sterile neutrinos \cite{PhysRevD.62.103007}, pseudo-Dirac neutrinos \cite{PhysRevLett.92.011101,PhysRevD.81.013006}, Lorentz or $CPT$ violation \cite{PhysRevD.72.065009}, and quantum gravity-induced decoherence \cite{PhysRevD.72.065019}.  Measuring the flavor ratio of astrophysical neutrinos is interesting both as a probe of the source of high energy cosmic rays and a test of fundamental particle physics.

\textit{Neutrino interactions in IceCube---}The IceCube detector consists of 5,160 digital optical modules (DOMs) instrumenting $1\rm{\,km}^{3}$ of clear ice at the South Pole \cite{2010RScI...81h1101H,icrev}.  Each DOM contains a photomultiplier and digitizing electronics that detect Cherenkov light emitted from secondary particles produced in neutrino interactions \cite{DAQPaper}.  Neutrino events are generally classified into two topologies: track-like, where the path of an outgoing charged particle is visible, and shower-like, where the region of light emission is too small to be resolved and appears point-like.  For both topologies the energy deposited within the detector can be reconstructed within $\sim15\%$ above $10\,\rm{TeV}$ \cite{energy_reco}.  Neutrino direction can be reconstructed with a median angular error of $\lesssim1^{\circ}$ for tracks versus $\sim15^{\circ}$ for showers above $100\,\rm{TeV}$ \cite{energy_reco}.  

In charged-current (CC) interactions, a neutrino deposits its energy into a charged lepton and a hadronic shower, and the topology of an event depends on flavor.  For $\nu_{e}$ CC interactions, the outgoing electron initiates an electromagnetic shower indistinguishable from the accompanying hadronic shower.  For $\nu_{\mu}$ CC interactions, the outgoing muon leaves a long track in addition to a hadronic shower.  If the muon leaves the detector, the deposited energy is only a lower bound on the neutrino energy.  For $\nu_{\tau}$ CC interactions, the outgoing tau decays quickly and is difficult to resolve for energy $\lesssim1\;\mathrm{PeV}$ \cite{Learned:1994wg,PhysRevD.86.022005,energy_reco}.  However, tracks may be observed from the muonic decay of the tau with $17.4\%$ branching ratio.  In neutral-current (NC) interactions of all flavors, a neutrino deposits on average $\sim1/3$ of its energy into a hadronic shower but with a cross section $\sim1/3$ of the CC cross section \cite{nuint}.  Above $\sim 10\,\mathrm{TeV}$, neutrino fluxes become attenuated by interactions in the Earth, though $\nu_\tau$ fluxes are regenerated by subsequent tau decay to neutrinos \cite{1998PhRvL..81.4305H}. 

The backgrounds in astrophysical neutrino searches are muons and neutrinos produced by cosmic ray air showers in Earth's atmosphere. Muons dominate the trigger rate in IceCube and usually create long tracks.  However, they can also appear shower-like if they undergo a single catastrophic energy loss inside the detector.  Atmospheric neutrinos are usually divided into two categories.  The first arises from the decay of kaons, pions and muons, producing mostly $\nu_\mu$ \cite{1983PhRvL..51..223G,2006JHEP...10..075G,2012PhRvD..86k4024F}.  Since time dilation causes decay to be less likely than interaction at high energy, the neutrino energy spectrum is asymptotically one power steeper than the primary cosmic-ray spectrum, and the angular distribution is peaked at the horizon.  Time dilation also suppresses $\nu_e$ from muon decay down to the detector depth, and the remaining $\nu_e$ are from kaon decays at the level of $\nu_e/\nu_\mu\approx 4$\%.  The flux of atmospheric $\nu_e$ has recently been measured in the TeV range by IceCube \cite{dc_cascades}.

The second category, yet to be observed, results from the prompt decay of charm mesons \cite{1978PhLB...78..635B,1983ICRC....7...22V,1999PhLB..462..211V,2000PhRvD..61e6011G,2003AcPPB..34.3273M,2008PhRvD..78d3005E}, yielding a nearly equal mixture of $\nu_e$ and $\nu_\mu$, but negligibly small $\nu_\tau$ \cite{PhysRevD.59.093003}.  Henceforth referred to as the charm neutrino flux, it should follow the same $\sim E^{-2.7}$ spectrum as primary cosmic rays and also be isotropic.  Since atmospheric backgrounds from muons and $\nu_\mu$ produced in $\pi/K$ decay are largely track-like, astrophysical events dominate over backgrounds down to lower energies in the shower channel, and a contribution from charm decay may be more easily identified \cite{2004JCAP...11..009B}.

\begin{figure}
\includegraphics{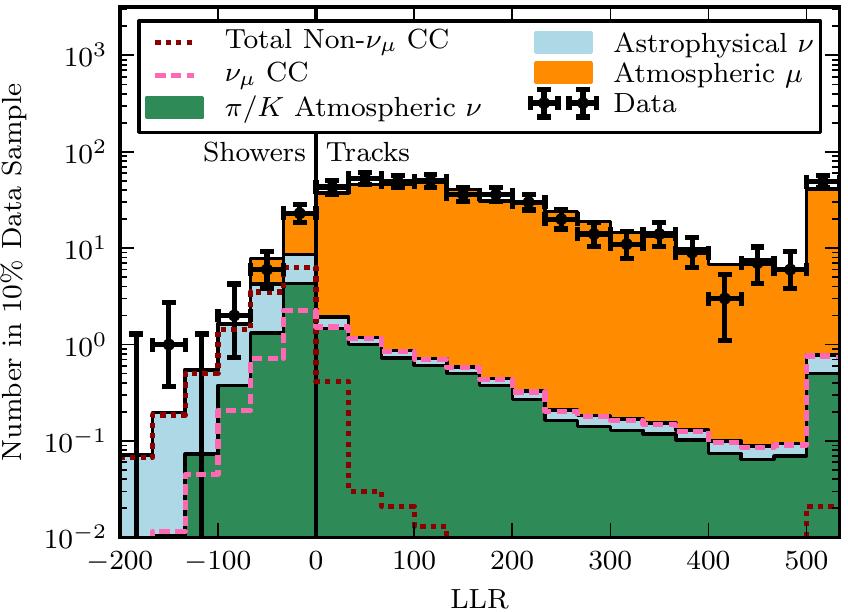}
\caption{The log likelihood ratio between shower and track reconstructions for veto-passing events with more than 1500 photoelectrons.  Error bars are $68\%$ Feldman-Cousins intervals.  The contribution of muons is determined from a muon control sample.  The dotted lines show the total amount of $\nu_\mu$ CC events (pink) and all non-$\nu_\mu$ CC events (maroon) from the best-fit distributions of astrophysical and $\pi/K$ neutrinos.  The last bin contains all overflow events with $\mathrm{LLR}>500$.}
\label{fig:deltallh}
\end{figure}

\textit{Event selection---}Data collected at IceCube in 974 days from May 2010 to May 2013 are used in this analysis.  During the design of the event selection criteria, $90\%$ of data was kept blind.  Following the same strategy as the previous 3-year analysis \cite{hese_paper,hese2}, events are selected using an outer layer of DOMs to veto the vast majority of incoming muons, isolating neutrino interactions of all types starting within the detector from across the entire sky. Also similarly, the muon background rate is estimated with a control sample.  Using outer DOMs to tag incoming muons, an inner volume geometrically similar to the full fiducial volume is defined with its own veto layer of DOMs.  After correcting for its smaller fiducial volume, the rate of tagged but unvetoed events yields the muon background rate in the full detector.

Down-going atmospheric neutrinos can also be vetoed by accompanying muons from the same cosmic ray air shower that reach the detector \cite{atmonu_veto}.  The veto probability is determined using the analytic calculation described in \cite{newvetopaper}, accounting for muons from both the same decay as the neutrino and other decays in the air shower.  The resulting suppression of down-going atmospheric neutrino events distinguishes them from the isotropic distribution expected from a diffuse astrophysical flux.  For the charm neutrino flux, otherwise isotropic, this suppression is the only distinguishing feature if the astrophysical flux has a power-law index close to 2.7 and a non-standard flavor ratio $(1:1:0)_\oplus$.

Showers and tracks are classified by performing a per-event maximum likelihood analysis of the first photon arrival times in every DOM.  Each event is reconstructed according to the hypothesis of an infinite track with constant light emission along its path \cite{2004NIMPA.524..169A} and the hypothesis of a point-like shower \cite{2011APh....34..420A}, yielding likelihoods $L_{\mathrm{Track}}$ and $L_{\mathrm{Shower}}$.  A log likelihood ratio, $\mathrm{LLR}=-\ln(L_{\mathrm{Shower}}/L_{\mathrm{Track}})$, is formed, with negative values being considered showers and positive values tracks.

Figure~\ref{fig:deltallh} shows a distribution of $\mathrm{LLR}$ for veto-passing events in the $10\%$ unblind data sample producing more than 1500 total photoelectrons (PE). Best-fit neutrino distributions for a $(1:1:1)_\oplus$ composition (discussed later) are shown, and the prediction for muons is obtained from the corresponding likelihood ratio distribution in the muon control sample.  The agreement with data illustrates that the control sample reliably predicts the rates of both shower-like and track-like muons.  Also shown is the combined distribution of astrophysical and atmospheric $\nu_{\mu}$ CC events, illustrating that most are classified as tracks.  The remaining $30\%$ of $\nu_\mu$ CC events classified as showers arise when the outgoing muon has too little energy to be resolved or escapes near the edge of the detector.

In the final selection, shower-like events above 1500 PE are selected, while for tracks, only events above 6000 PE are selected due to the larger background from penetrating muons.  The deposited energy and direction of each event is reconstructed using the full timing distribution of recorded photoelectrons in every DOM. For events classified as showers, point-like light emission is assumed, whereas for tracks, the energy deposition is unfolded along the path of the track \cite{energy_reco}.  To avoid systematic uncertainties relating to muons, down-going, shower-like events below $20\;\mathrm{TeV}$ are excluded because they are dominated by muons.

\begin{figure*}
\includegraphics{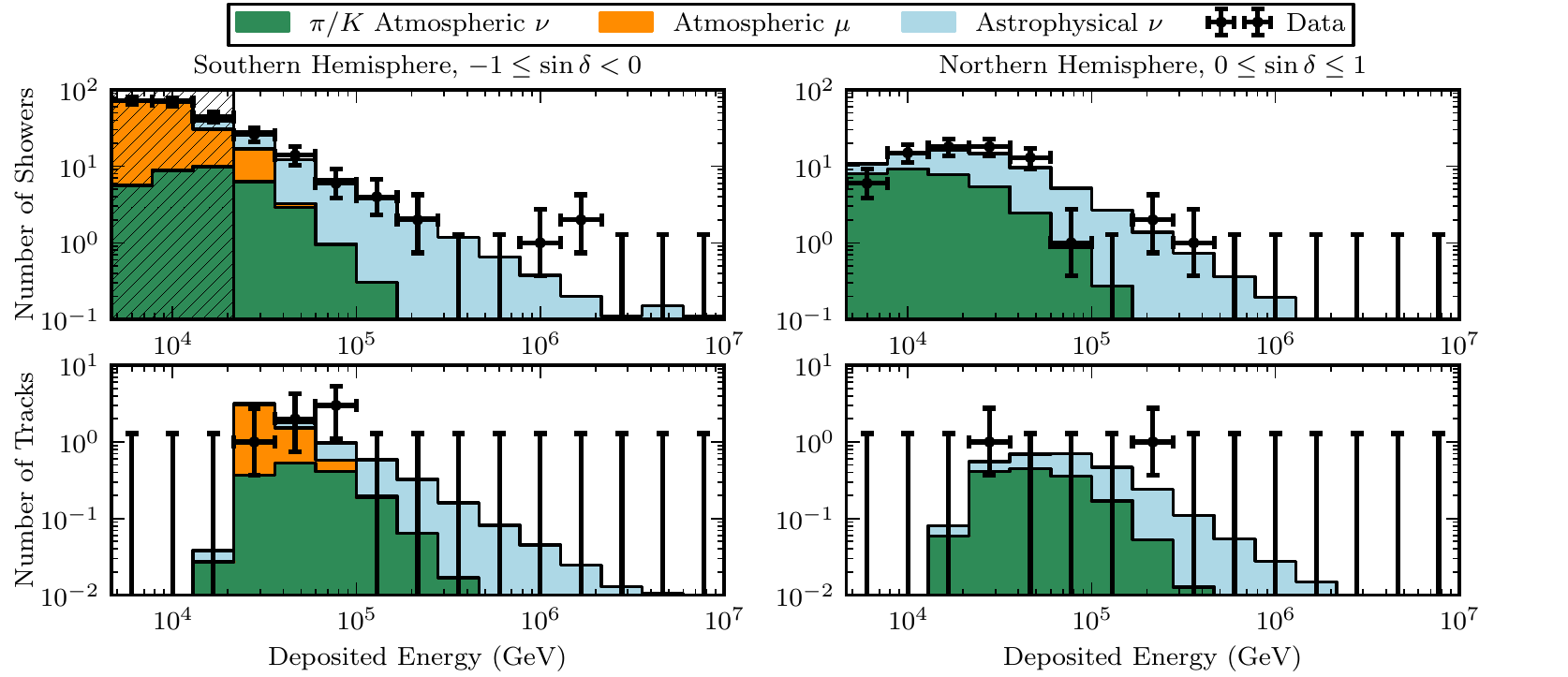}
\caption{The best-fit deposited energy distributions for showers and tracks, divided into the southern (down-going) and northern (up-going) samples, assuming a power-law astrophysical flux with $(1:1:1)_\oplus$ flavor ratio at Earth.  Showers in the southern sky below $20\;\mathrm{TeV}$ are dominated by muons and excluded; however, the prediction from the control sample measurement is shown in this region. Though not shown, 4 bins in declination ($\delta$) are used in the fit.}
\label{fig:energy}
\end{figure*}

\textit{Statistical Analysis---}To measure the flavor ratio of the astrophysical flux, we follow the approach of earlier IceCube analyses \cite{ic59_muons} and perform a binned, maximum likelihood fit over the 2D distributions of deposited energy and reconstructed declination of both showers and tracks.  The expected count in each bin is calculated from Monte Carlo simulation and depends on a set of nuisance parameters, which describe systematic uncertainties, and physics parameters, which are of interest to be measured.  The likelihood contains two terms---the first describing the Poisson distribution of bin counts and the second penalizing deviations of nuisance parameters from their central values according to their uncertainty \cite{pdg}.  To construct confidence intervals and perform hypothesis tests, we use the profile likelihood method \cite{profile_llh} and minimize the likelihood over nuisance parameters.  Unless noted, we assume that profile likelihood ratios follow a $\chi^2$ distribution \cite{wilks1938}. 

Systematic uncertainties are either detector-related, such as DOM optical efficiency and ice optical properties, or theoretical, such as neutrino fluxes and cross sections.  All tend to uniformly scale the rates of tracks and showers, maintaining their ratio and leaving their energy and angular distributions unaffected.  Thus, nuisance parameters describing backgrounds become overall rate scaling factors applied to the distributions of $\pi/K$ neutrinos, charm neutrinos, and muons.  To describe atmospheric neutrinos, we use the flux calculation of HKKMS \cite{Honda2006} for $\pi/K$ neutrinos and ERS \cite{EnbergPrompt} for charm neutrinos.  A correction is included to describe the cosmic ray knee in the model of \cite{GaisserImprovedCRSpectrum}, and the HKKMS flux is extrapolated above 10 TeV as described in \cite{ic59_muons}.  No priors are used to constrain the scalings of these atmospheric neutrino distributions.  For muons, although the control sample constrains the overall passing rate, there are insufficient statistics in its energy and angular distribution, so simulation is used.  It is based on a parametrization of the deep-ice muon flux \cite{improvedveto,jvsthesis} obtained from CORSIKA air-shower simulation \cite{CORSIKA} and the cosmic ray model of \cite{GaisserImprovedCRSpectrum}.  Since muons are the dominant background for astrophysical tracks, we allow the scalings applied to track-like muons and shower-like muons to float independently, accounting for uncertainties such as ice properties, energy loss cross sections, and muon bundle multiplicity that could skew the ratio of tracks to showers.  These scalings are, however, constrained by a Gaussian prior of $8.4\pm4.2$ events each, derived from the 4 surviving events each in the track-like and shower-like muon control samples.

When placing limits on the flavor ratio, nuisance parameters also include those describing the astrophysical flux.  For this analysis, we use an isotropic, power-law flux with the following parametrization for each neutrino flavor,
\begin{equation}
\Phi_{\alpha}(E)  = 3\Phi_{0}f_{\alpha,\oplus} \left(\frac{E}{100\;\mathrm{TeV}}\right)^{-\gamma}\rm{,}
\label{eq:flux}
\end{equation}
where $f_{\alpha,\oplus}$ is the fraction of each flavor at Earth, $\gamma$ is the spectral index, and $\Phi_0$ is the average flux of $\nu$ and $\bar{\nu}$ per flavor at $100 \rm{\,TeV}$.  An equal $\nu$ and $\bar{\nu}$ flux is assumed.  Though this does not hold for neutrinos of photohadronic origin, there are consequences for a flavor ratio measurement only from yet-unobserved $\bar{\nu}_e$ interactions at the $6.3 \rm{\,PeV}$ Glashow resonance \cite{PhysRevD.84.033006,1475-7516-2011-10-017}, too high in energy to have a significant impact with currently available statistics.


\textit{Results---}After all selection criteria, 129 showers and 8 tracks remain in the final event sample, forming a superset of the earlier 3-year sample with 28 showers and 8 tracks \cite{hese2}.  Before attempting to constrain the astrophysical flavor ratio, it is necessary to verify that the adopted isotropic, power-law model of the astrophysical flux adequately describes the data.  Assuming a flavor composition of $(1:1:1)_{\oplus}$ at Earth, the best-fit distributions are shown in Fig.~\ref{fig:energy}.  Noteworthy are the best-fit astrophysical flux parameters, $\gamma=2.6\pm0.15$ and $\Phi_0 = \left(2.3\pm0.4\right) \times 10^{-18}\,\mathrm{GeV}^{-1}\,\mathrm{s}^{-1}\,\mathrm{cm}^{-2}\,\mathrm{sr}^{-1}$.  While being compatible with the previous 3-year result \cite{hese2}, the spectral index is substantially different from $\gamma=2$, which is rejected at $3.0\sigma$.  The preference for $\gamma>2$ comes mostly from low-energy data rather than a lack of events above several $\mathrm{PeV}$.  A high-energy cutoff in the astrophysical flux of the form $\propto E^{-2}\exp{(-E/E_c)}$ is also disfavored with respect to a power law at $2.9\sigma$.  Finally, the power-law model with $(1:1:1)_\oplus$ flavor ratio is in agreement with data with a goodness-of-fit p-value of $0.13$.

A large charm flux to explain low-energy data is disfavored since the suppression of down-going events by accompanying muons is not observed, and the best-fit scaling of the ERS flux is 0.  Even fixing the ERS scaling at its $90\%$ upper limit of 3.4 obtained here, the astrophysical index only changes to $\gamma=2.5$.  These results are consistent with a recent, dedicated IceCube measurement of the astrophysical spectral index and charm flux with improved veto techniques \cite{improvedveto}.   Nuisance parameters describing $\pi/K$ neutrinos and muons are also consistent with expectations from the HKKMS flux and the control sample measurement.

This analysis is sensitive to the astrophysical flux in the neutrino energy range $35\,\mathrm{TeV}$ -- $1.9\,\mathrm{PeV}$.  The lower and upper bounds of this range, $E_{\mathrm{low}}$ and $E_{\mathrm{up}}$, were calculated separately by fixing the astrophysical spectral index and normalization at their best-fit values, excluding the flux with $E < E_{\mathrm{low}}$ or $E > E_{\mathrm{up}}$, respectively, refitting the data with nuisance parameters left free, and finding the values for $E_{\mathrm{low}}$ or $E_{\mathrm{up}}$ that decreased the log likelihood by $1/2$ each.

\begin{figure}
\includegraphics{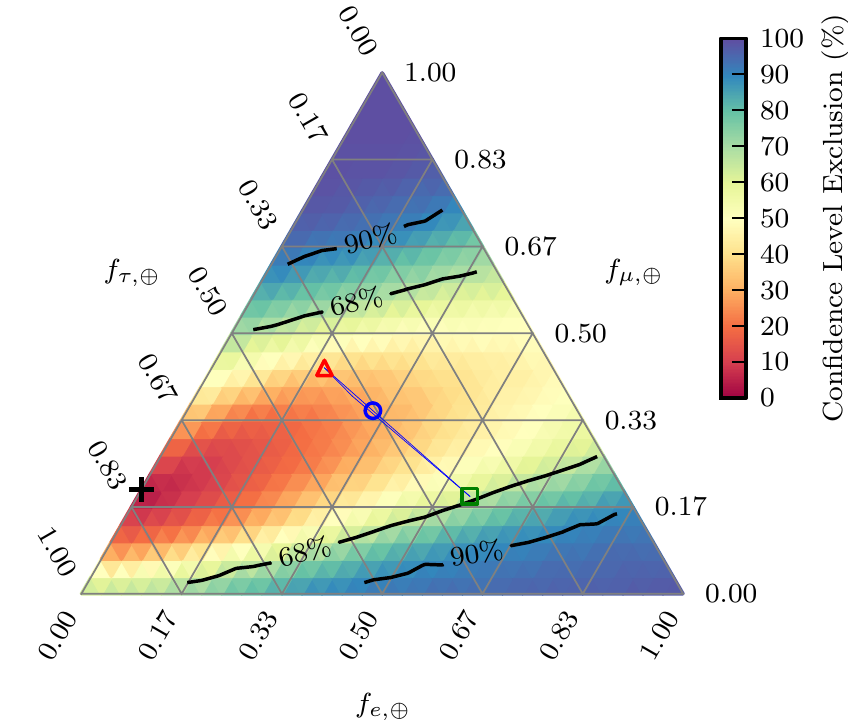}
\caption{The exclusion regions for astrophysical flavor ratios $(f_{e}:f_{\mu}:f_{\tau})_\oplus$ at Earth.  The labels for each flavor refer to the correspondingly tilted lines of the triangle.  Averaged neutrino oscillations map the flavor ratio at sources to points within the extremely narrow blue triangle.  The  $\approx(1:1:1)_\oplus$ composition at Earth, resulting from a $(1:2:0)_S$ source composition, is marked with a blue circle.  The compositions at Earth resulting from source compositions of $(0:1:0)_S$ and $(1:0:0)_S$ are marked with a red triangle and green square, respectively.  Though the best-fit composition at Earth (black cross) is $(0:0.2:0.8)_\oplus$, the limits are consistent with all compositions possible under averaged oscillations.}
\label{fig:flav}
\end{figure}

With a power-law astrophysical flux describing the data, we then further allow the flavor composition at Earth to float and calculate exclusion regions according to the Feldman and Cousins approach \cite{feldmancousins}, as shown in Fig.~\ref{fig:flav}.  Though the best-fit composition is $(0:0.2:0.8)_\oplus$ at Earth, the limits are compatible with all standard flavor compositions possible under averaged neutrino oscillations at $<68\%$ confidence level. 

With showers and tracks serving as the only two identifiers for three flavors in this analysis, there is an inherent degeneracy in the determination of astrophysical flavor ratios.  This is reflected in the strong anti-correlation between $\nu_e$ and $\nu_{\tau}$ components, which both produce mostly showers.  The degeneracy is broken mainly by two effects---the shift in the $\nu_\tau$ deposited energy distribution caused by invisible energy lost to neutrinos in tau decay and the lack of observed $\bar{\nu}_e$ Glashow resonance events.   The preference for a $\nu_\tau$-like signature is not statistically significant, and future work to identify $\nu_{\tau}$ signatures at $\mathrm{PeV}$ energies may resolve this degeneracy.  

Since compositions produced by averaged neutrino oscillations (narrow blue triangle in Fig.~\ref{fig:flav}) are nearly orthogonal to the flavor degeneracy in IceCube, constraints on source flavor composition are possible but not yet significant.  After restricting to flavor ratios allowed by averaged neutrino oscillations, no source composition can be excluded at $>68\%$ confidence level, and this remains true even with the additional constraint $f_{\tau,S}=0$ expected at astrophysical sources.  

Having found agreement with the predictions of averaged neutrino oscillations, constraints are placed on non-standard flavor compositions producing a large $\nu_e$ or $\nu_\mu$ fraction at Earth.  A maximally track-like, pure $\nu_\mu$ signature of $(0:1:0)_\oplus$ is excluded at $3.3\sigma$ and a purely shower-like $\nu_e$ signature of $(1:0:0)_\oplus$ at $2.3\sigma$.

These results contrast with an earlier analysis of IceCube's 3-year data, which found a preference for $(1:0:0)_\oplus$ over $(1:1:1)_\oplus$ at $92\%$ confidence level \cite{heseflav}.  We attribute this discrepancy mainly to two unaccounted for effects --- partial classification of $\nu_\mu$ CC events as showers and systematic uncertainty on muon background.  Repeating their analysis but accounting for the $\sim30\%$ of $\nu_\mu$ CC events classified as showers and using a profile likelihood incorporating the $50\%$ uncertainty in muon background, a $(1:0:0)_\oplus$ best-fit is still obtained but neither $(1:1:1)_\oplus$ or our best-fit of $(0:0.2:0.8)_\oplus$ are excluded at $>68\%$ confidence level.  Since only shower and track counts were analyzed, the tighter constraints reported here result from the use of energy and directional information in addition to the lower energy data.

Future measurements of the flavor ratio at IceCube will use improved veto techniques, include up-going tracks starting outside the detector, and search for high-energy signatures of $\nu_\tau$.  With these improvements, measuring the flavor composition at astrophysical sources and precision tests of neutrino oscillations over astronomical distances will be in reach.

\begin{acknowledgments}

We acknowledge the support from the following agencies:
U.S. National Science Foundation-Office of Polar Programs,
U.S. National Science Foundation-Physics Division,
University of Wisconsin Alumni Research Foundation,
the Grid Laboratory Of Wisconsin (GLOW) grid infrastructure at the University of Wisconsin - Madison, the Open Science Grid (OSG) grid infrastructure;
U.S. Department of Energy, and National Energy Research Scientific Computing Center,
the Louisiana Optical Network Initiative (LONI) grid computing resources;
Natural Sciences and Engineering Research Council of Canada,
WestGrid and Compute/Calcul Canada;
Swedish Research Council,
Swedish Polar Research Secretariat,
Swedish National Infrastructure for Computing (SNIC),
and Knut and Alice Wallenberg Foundation, Sweden;
German Ministry for Education and Research (BMBF),
Deutsche Forschungsgemeinschaft (DFG),
Helmholtz Alliance for Astroparticle Physics (HAP),
Research Department of Plasmas with Complex Interactions (Bochum), Germany;
Fund for Scientific Research (FNRS-FWO),
FWO Odysseus programme,
Flanders Institute to encourage scientific and technological research in industry (IWT),
Belgian Federal Science Policy Office (Belspo);
University of Oxford, United Kingdom;
Marsden Fund, New Zealand;
Australian Research Council;
Japan Society for Promotion of Science (JSPS);
the Swiss National Science Foundation (SNSF), Switzerland;
National Research Foundation of Korea (NRF);
Danish National Research Foundation, Denmark (DNRF)

\end{acknowledgments}

%



\end{document}